# *The age of spiritual machines*: Language quietus induces synthetic altered states of consciousness in artificial intelligence

Jeremy I Skipper[1*], Joanna Kuc[1], Greg Cooper[3], & Christopher Timmermann[4]

[1] Experimental Psychology, University College London, UK
[2] Clinical, Educational and Health Psychology, University College London, UK
[3] Centre for Psychedelic Research, Department of Brain Sciences, Imperial College London, London UK
[*] Corresponding author: jeremy.skipper@ucl.ac.uk

## Abstract

How is language related to consciousness? Language functions to categorise perceptual experiences (e.g., labelling interoceptive states as 'happy') and higher-level constructs (e.g., using 'I' to represent the narrative self). Psychedelic use and meditation might be described as altered states that impair or intentionally modify the capacity for linguistic categorisation. For example, psychedelic phenomenology is often characterised by 'oceanic boundlessness' or 'unity' and 'ego dissolution', which might be expected of a system unburdened by entrenched language categories. If language breakdown plays a role in producing such altered behaviour, multimodal artificial intelligence might align more with these phenomenological descriptions when attention is shifted away from language. We tested this hypothesis by comparing the semantic embedding spaces from simulated altered states after manipulating attentional weights in CLIP and FLAVA models to embedding spaces from altered states questionnaires before manipulation. Compared to random text and various other altered states including anxiety, models were more aligned with disembodied, ego-less, spiritual, and unitive states, as well as minimal phenomenal experiences, with decreased attention to language and vision. Reduced attention to language was associated with distinct linguistic patterns and blurred embeddings within and, especially, across semantic categories (e.g., 'giraffes' become more like 'bananas'). These results lend support to the role of language categorisation in the phenomenology of altered states of consciousness, like those experienced with high doses of psychedelics or concentration meditation, states that often lead to improved mental health and wellbeing.

*I feel one with everything around me. My 'self' or ego dissolves into nothingness.*

From the 'Ego Dissolution Scale'[1]

*My brain chatter began to disintegrate... my consciousness soared into an all-knowingness, a 'being at one' with the universe... I no longer perceived myself as a whole object separate from everything... I retained only a vague idea of who I was...*

JB Taylor[2]

## Introduction

The Ego Dissolution Scale (EDS) has two subfactors, 'unity' and 'ego-loss', states that are often reported after high doses of psychedelics and by experienced meditators[1]. Given her description in the epigraph, one might assume JB Taylor was either a frequent psychedelic user or an advanced meditator. However, this quote is from Dr Taylor's book, *My Stroke of Insight,* in which she describes the experience of a cerebrovascular event that led to global aphasia, an inability to speak or understand language[2]. There are many similar anecdotes[1] of people deprived of or having lost language, describing their experience upon acquiring or recovering language[4]. These suggest that language plays a mechanistic and causal role in determining the phenomenology of more 'typical' and altered states of consciousness[4,5].

Indeed, the empirical evidence for this position is amassing. Behavioural studies suggest that words facilitate the conscious experience of non-linguistic stimuli, through categorically organising relevant sensory information[6]. For example, colour, motion, and object words promote the detection of visual colour, motion, and objects, even bringing them into consciousness[4]. This is supported by neurobiological data, with damage and aphasia, split-brain patients, intracarotid anaesthetic, and disorders of consciousness all suggesting language is causally related to consciousness[4]. For example, injection of an anaesthetic into the left or right intracarotid artery is more likely to result in unconsciousness in individuals more left or right lateralised for language functioning, respectively[7].

Further supporting the link between language and consciousness, numerous studies have shown a similar relationship between language and psychedelic-induced altered states. Evidence suggests that these experiences often involve ineffability—an inability to adequately convey experiences in words[8–10]. This might manifest as slower, reduced, and less complex language output[11–14]. When language is engaged, it tends to be more unusual and bizarre, less predictable, and involves altered

---

[1] For example, in *A Stitch of Time: The Year a Brain Injury Changed My Language* and in other sources, Lauren Marks describes her aphasia as 'My entire environment felt interconnected, like cells in a large, breathing organism. ... I felt less like myself and more like everything around'[3]

semantic topics (e.g., a decreased focus on the self)[11,13,15–17]. Research has described an increased spread of semantic activation under psychedelics, corresponding to enhanced indirect priming, increased semantic distances, and more semantic errors[18–21]. Although these changes are often linked to creativity, they might also be described as a suppressed and malfunctioning language system.

Corroborating this behavioural data, numerous studies have demonstrated that psychedelics affect core language-related brain regions, most notably the posterior inferior frontal gyrus and the superior and middle temporal gyri[22–25]. These regions are particularly implicated in language processing and are also rich in 5-HT(Serotonin)$_{2A}$ receptors, claimed to be a primary target of psychedelics[26]. The high concentration of 5-HT$_{2A}$ receptors in these areas[27,28] aligns with the central role of language in the phenomenology of psychedelic unitive states and self-loss. Indeed, Cooper et al.[29] found that individual differences in the 'Ego Dissolution Inventory' (EDI[30]) scores under psychedelics like LSD or psilocybin are closely linked to variations in these same language-associated regions. Moreover, under psilocybin, there were marked changes in connectivity between language-related regions and visual and somatosensory cortices, indicating a decoupling of language-associated neural processes from sensory processing.

*Mechanisms*

How does language play a role in inducing altered states? We have proposed that the neurobiology of language is a whole brain process that connects words processed in auditory sensory regions to distributed regions of the brain involved in sensorimotor processing (e.g., auditory, visual, motor, somatosensory, and interoceptive regions) and memory (e.g., regions involved in autobiographical memory)[4,31]. Specifically, words flexibly organise our perceptual and more general experience of the world. This involves categorically organising external auditory and visual information (e.g., into colours, faces, and objects). Language also categorically organises internal information, like labelling lower-level interoceptive processing with emotional words, but also higher-level categories like our 'narrative self', largely involving stories constructed and manipulated with words. These categorical processes are established with learning and can become positively or negatively entrenched.

From this perspective, psychedelic phenomenology may be explained by the breakdown of the categorical functions of language[4,5]. That is, brain networks involved in language processing might cease to function typically and/or function at a reduced level or even be temporarily suspended altogether. Specifically, we suggest that auditory cortical activity decreases, diminishing or eradicating labels or word forms[32,33]. This occurs with hyperactivity or a decrease in inhibition in core frontal brain regions involved in semantic selection and retrieval[34–36]. This combination of processes in distributed language-related brain circuits disrupts entrenched linguistic categories, resulting in the reviewed changes to semantic representations. Thus, increased unity can be seen as the result of the information processed throughout the brain being less differentiated due to a lack of categorical linguistic constraints. Similarly, ego-loss would result from the diminished ability to maintain higher-level language categories like 'me'.

Similarly, certain features of minimal phenomenological experiences achieved by some during advanced mediation practices, and possibly during certain psychedelic states[37], can be parsimoniously explained by the temporary reorganisation and perhaps cessation of language. 'Pure awareness' is defined by the absence of the contents of consciousness, including narrative, and self-referential thoughts, offering an open, unbounded state of being[38]. The cognitive demands of language may act as a persistent mediator, drawing awareness away from this boundless state and into a realm dominated by lower-level (e.g., object) and higher-level categories like the 'I'-thoughts related to the 'narrative self'. Consequently, while language enables complex cognition and communication, it obstructs the 'Peace, Bliss, and Silence' and 'Emptiness and Non-egoic Self-awareness' factors of the Minimal Phenomenological Experience (MPE-92M) questionnaire[39]. Furthermore, that language is significantly dysregulated and that categorisation is unconstrained during acute altered states, aligns well with the notion of 'epistemic openness' associated with the minimal phenomenal experience[40].

## *Hypotheses*

We sought to test this theoretical framework by generating artificial alternate states of consciousness in machine intelligence. Indeed, our theory conceptually aligns well with multimodal deep neural networks like the CLIP (Contrastive Language-Image Pre-training) and FLAVA (Flow-based Language and Vision Attention) models. Both utilise shared semantic embedding spaces to relate text and image modalities. These models capture linguistic and sensory interactions akin to those in distributed language networks of the human brain. Just as we suggest that language categorically organises sensory, motor, and interoceptive experiences, CLIP and FLAVA map language content and visual input onto a shared space, mirroring the process by which we suggest language shapes our subjective experience. CLIP, trained on an extensive dataset of image-caption pairs, models how language forms and maintains perceptual categories. FLAVA, with its multimodal encoder, captures a more complex interplay between language and sensory processing. Crucially, these models possess manipulable text and image 'attention' parameters. By manipulating weights in these shared semantic embedding spaces, we might be able to simulate the altered language functions observed in psychedelic and meditative states, potentially leading to a simulacrum of unitive states, ego dissolution, and minimal phenomenal experiences.

Here, we developed an analytical approach to test these parallels between our proposed mechanistic account and artificial intelligence (AI). Specifically, CLIP and FLAVA models use attention mechanisms that we can manipulate to increase or decrease attention weights to text and/or images in various combinations. We first simulated altered states of consciousness using text descriptions (Figure 1A). We then evaluated their alignment with factors from validated altered state questionnaires, such as the unity factor of the Altered States of Consciousness Rating Scale (ASC)[41], the ego-loss factor of the EDS[1], and minimal phenomenal experience (MPE) as measured by the MPE-92M questionnaire[39]. By manipulating the attention mechanisms of the models, we assessed how these alterations influenced the alignment between the simulated states ('responses') and individual questions from specific questionnaire factors ('prompts'). Similar to the effects of psychedelics on language, we expected deviations from baseline attention to degrade performance, measured by the cosine similarity between the shared embedding space representation of the prompt before manipulation and the response after attentional manipulation (Figure 1B). However, we anticipated degradation would be relatively spared, depending on the factors from which the prompt-response pairs originate and the particular text-image weight combinations.

Specifically, we hypothesised that reducing attention to text and images would generally degrade performance (measured by prompt-response similarity), but would be relatively preserved for unity, ego-loss, and MPE factors. In contrast, factors such as anxiety and cognition that rely more heavily on linguistic processing would show greater degradation. Conversely, unity, ego-loss, and MPE prompts would show greater degradation when attention to text is high whereas anxiety and cognition would be more well preserved. These hypothesised changes were expected to be reflected in the type of language in the simulated altered states, e.g., reduced self-reference under low text attention, mimicking ego-loss and increased self-referential content with high text attention. Consistent with our mechanistic account, we hypothesised that these relative effects of reduced attention to text are caused by the blurring of linguistic categories (generating unity, ego-loss, and MPE). We tested this by examining within and across-category similarity for the same text and image weights we used to test other hypotheses, expecting low attention to text to result in the most cross-category similarity.

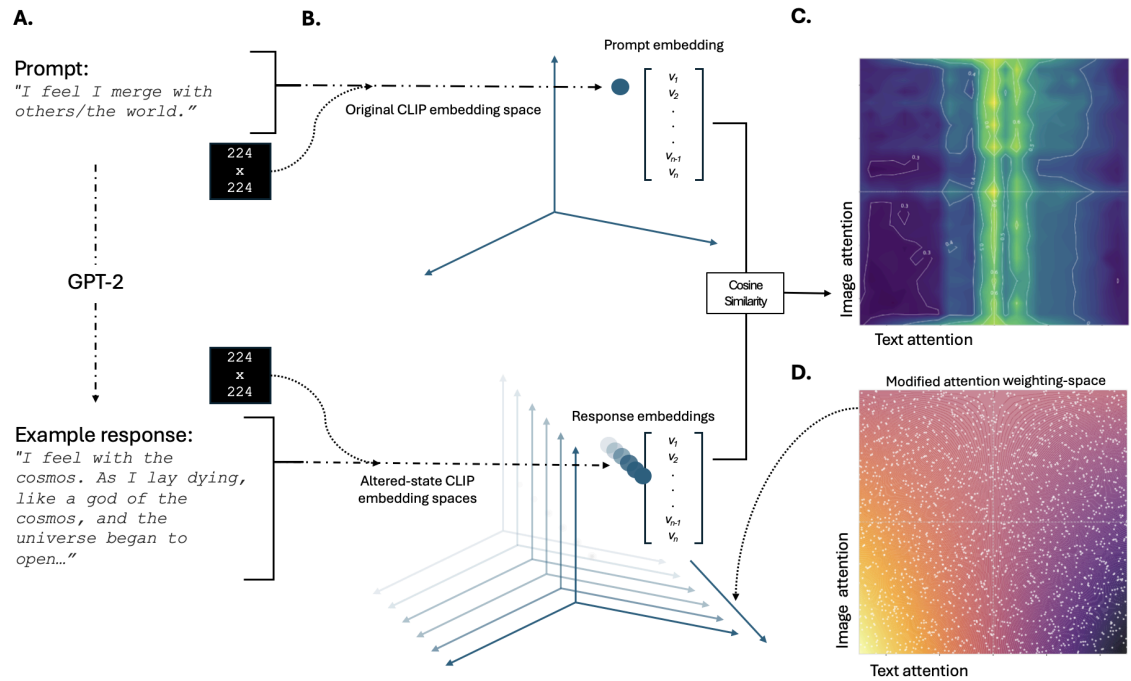

***Figure 1***. *Pipeline for generating similarity scores across the attentional weight space. A. Prompts from altered states questionnaires are provided to GPT-2 to generate 75-token responses. B. Prompts and a plain black image (256x256 pixels) are used to obtain embeddings from unmodified multimodal models (CLIP or FLAVA). Embeddings for random GPT-2 tokens (not shown) alongside embeddings for responses and the black image are then retrieved from 2013 models with systematically modified text and image attention weights. C. Cosine similarity is calculated between prompt and response encodings, between prompts and random samples, and averaged across questionnaire factors to produce similarity scores for the text-image attentional weight space. D. Text and image attention weight pairings for each of the 2013 models were generated via Latin hypercube sampling of a text and image weight space, where white points indicate sampled locations.*

## Results

To evaluate our proposal that language quietus induces synthetic altered states of consciousness, we focused on comparing factors from questionnaire-based metrics of altered states. We emphasised contrasts between the Anxiety and Unity factors of the ASC, as they appear to represent opposing experiences. We first demonstrate, qualitatively and quantitatively, that these factors produce distinct degradation patterns in models as a function of attention to text and images. We then applied searchlight analyses to precisely map how these attentional manipulations affected different factors. Finally, we analysed the associated language features and examined category coherence within and across states to further understand these effects.

## *Similarity*

Figure 2 visualises the relationship between prompt and response cosine similarity scores as a function of text and image attentional weights. It shows that similarity does not degrade uniformly; instead, the largest reduction in similarity occurs with negative text and image weights (i.e., quadrant one or Q1) across all ASC factors (Figure 2, left column). This pattern varies by specific altered state factor, though it remains relatively consistent across different models. Comparing the top and bottom panels of the second and third columns of Figure 2, both the CLIP and FLAVA models show that the Anxiety factor of the ASC differs from the Unity factor in all quadrants, in a relatively similar manner across models. Factors from other altered state questionnaires show subtle but distinctly different patterns (Figure 2, last column).

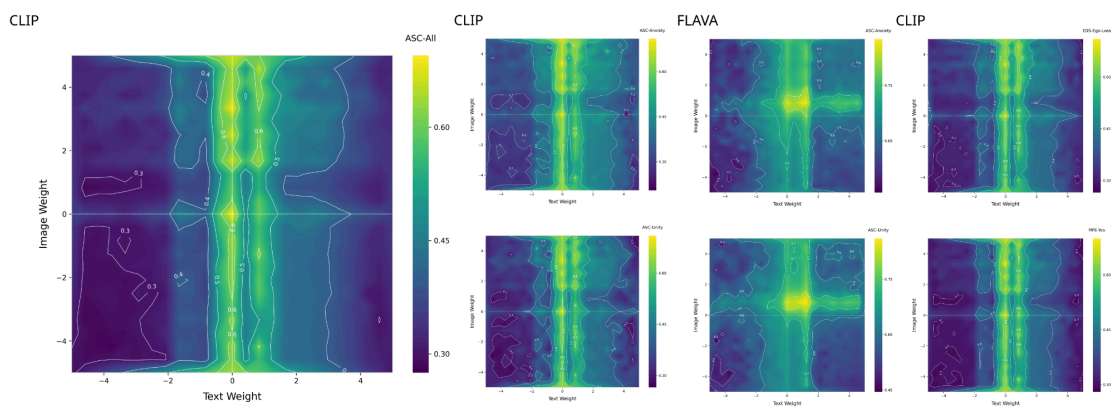

*Figure 2.* Heatmaps of prompt and response cosine similarity scores as a function of text (x-axis) and image (y-axis) attention weights across factors and models. The first panel shows the combined 11 Altered States of Consciousness Rating Scale (ASC) factors. The second and third columns present ASC Anxiety (top) and Unity (bottom) factors for the CLIP and FLAVA models. The fourth column depicts the Ego Dissolution Scale (EDS) Ego-Loss factor (top) and Minimal Phenomenal Experience (MPE) as measured by the MPE-92M questionnaire (bottom) for the CLIP model. Cooler colours indicate lower (blue) and warmer colours higher (yellow) similarity scores. White contour lines mark regions with similar response values, highlighting gradients within the attentional weight space (Figure 1).

To qualitatively visualise the differential degradation of similarity, we found the medium difference between prompt and random cosine similarity scores for each of the 11 ASC factors in each of the four quadrants (Figure 3). Overall, different quadrants grossly load differently on different ASC factors, with Q1 and Q2 showing the largest variations.

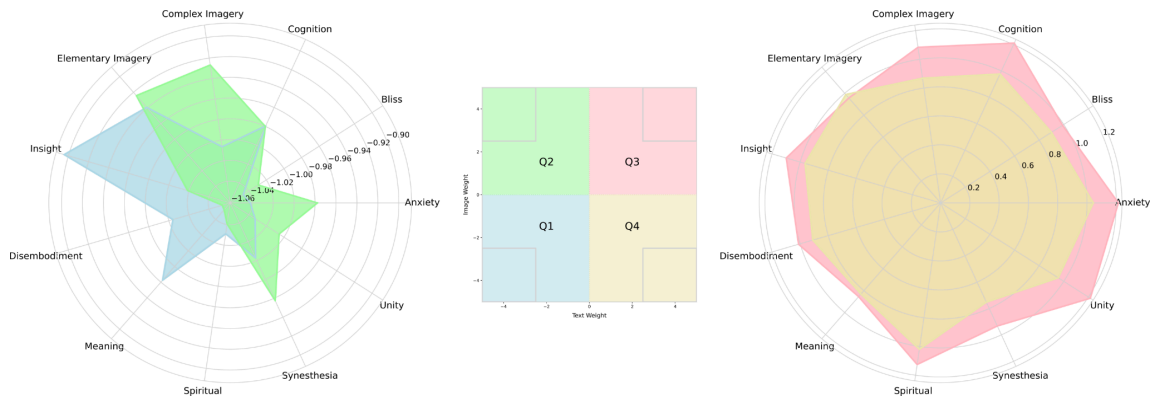

*Figure 3.* Spider plots showing cosine similarity scores for the 11 ASC factors across four quadrants of the 2D attentional weight space, defined by text and image attention levels. Each axis represents one ASC factor, displaying median differences between prompt-response and random response scores, z-normalised. The quadrants (Q1–Q4) in the legend (centre) represent distinct combinations of high and low attentional weights, with colours corresponding across panels.

## *GAM*

We next sought to quantify the observed variable relationships between text and image attention weights and cosine similarity scores for the CLIP and FLAVA models as they related to different altered states. The results of Generalised Additive Model (GAM) analyses are presented in Figure 4.

**CLIP**

The top row of Figure 4 presents the GAM model results for CLIP, comparing all ASC factor scores to the random score. The model showed a strong fit, with a pseudo-R-squared value of 0.9099, indicating that 90.99% of the variance in the cosine similarity scores was explained by the model. Both text weight ($s(0)$) and image weight ($s(1)$) showed significant non-linear effects ($p < 1.11 \times 10^{-16}$), highlighting the complex relationships between these weights and the similarity scores. The main effect of Score Type ($f2$)) was also significant ($p < 1.11 \times 10^{-16}$), indicating an overall difference between ASC and random scores. Moreover, the interaction effects between text weight and score type ($te(0,2)$) and image weight and score type ($te(1,2)$) were both significant ($p < 1.11 \times 10^{-16}$), suggesting that the differences between ASC and random scores depended on specific attentional weight conditions.

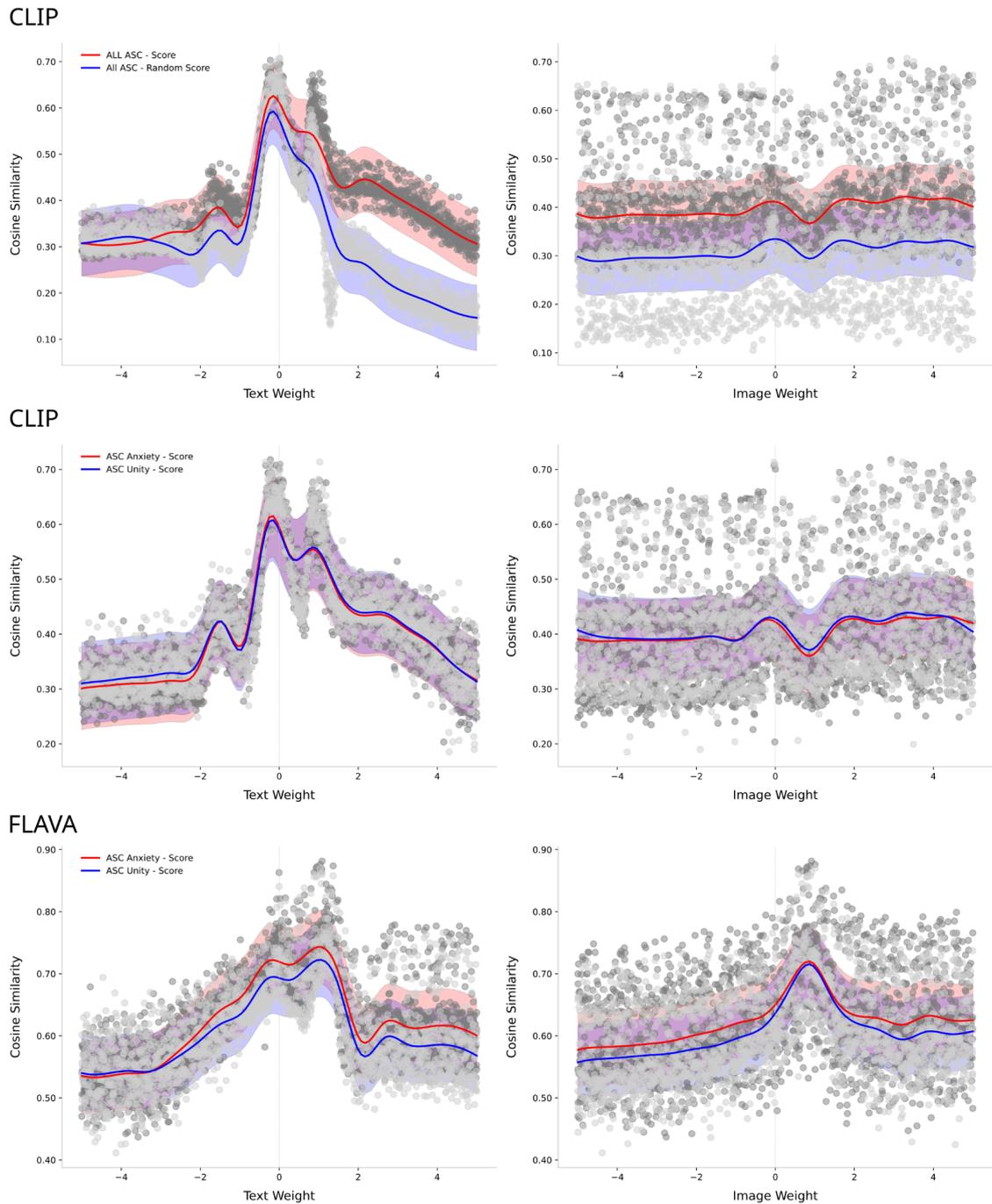

*Figure 4.* Scatter plots illustrate the relationship between text and image attention weights and cosine similarity scores for CLIP and FLAVA models. The top row displays the CLIP model results for all 11 ASC factors, comparing scores (red) and random scores (blue) for text weights (left) and image weights (right). The middle row shows the CLIP model results comparing ASC Anxiety (red) and ASC Unity factor (blue) scores, while the bottom row presents the same comparison for the FLAVA model. In all panels, grey circles represent cosine similarity scores for the 2,013 individual data points from the attentional weight space (Figure 1). The red and blue lines represent the nonlinear predictions generated by the Generalised Additive Model

*(GAM), capturing the relationship between attentional weights and cosine similarity scores. Shaded areas indicate 95% confidence intervals (CIs) for the GAM predictions.*

The middle row of Figure 4 shows the results of the GAM model for CLIP, comparing the ASC Anxiety and ASC Unity factors. The model fit was strong, with a pseudo-R-squared value of 0.8599, indicating that 85.99% of the variance in similarity scores was explained by the model. Both text weight ($s(0)$) and image weight ($s(1)$) had significant effects ($p < 1.11 \times 10^{-16}$), suggesting that the relationships between these weights and the similarity scores were non-linear. The main effect of Score Type ($f(2)$) was not significant ($p = 0.943$), suggesting that there was no overall difference between ASC Anxiety and Unity scores. However, the interaction effects between text weight and score type ($te(0,2)$) and image weight and score type ($te(1,2)$) were both significant ($p < 1.11 \times 10^{-16}$), indicating that the relationship between attentional weights and similarity scores differed between ASC Anxiety and ASC Unity factors for specific text-image weight pairs.

**FLAVA**

The bottom row of Figure 4 presents the GAM model results for FLAVA, comparing ASC Anxiety and ASC Unity factor scores. The model fit was strong, with a pseudo-R-squared value of 0.8536, indicating that 85.36% of the variance in the similarity scores was explained by the model. The text weight ($s(0)$) and image weight ($s(1)$) effects were both significant ($p < 1.11 \times 10^{-16}$), demonstrating non-linear relationships between attentional weights and similarity scores. The main effect of Score Type ($f(2)$) was also significant ($p = 7.46 \times 10^{-7}$), indicating an overall difference between ASC Anxiety and Unity factor scores. In addition, both interaction effects ($te0,2$) and $te(1,2)$ were significant ($p < 1.11 \times 10^{-16}$), suggesting that the differences between ASC Anxiety and Unity factors depended on specific attentional weight values.

## *Searchlight*

To quantify the significant non-uniform effects of text and image attention within the 2D attentional weight space, we developed a novel searchlight method. This approach systematically identifies localised effects of attentional manipulations using a sliding window, correcting for multiple comparisons, and clustering significant regions. This allowed us to pinpoint areas where attentional weights had the greatest impact on embedding spaces for responses simulating altered states. For example, negative text and reduced image attention may be linked to unity rather than anxiety states, as seen qualitatively and in the quantitative GAM analysis (Figures 2-4).

Indeed, the searchlight analysis reveals differential attentional weight effects across various ASC factors (Figure 5). The Imagery-related factors show distinct patterns. Elementary and Complex

Imagery tend to load more on regions with positive image weights (Q2 and Q3). However, when directly contrasted, Elementary Imagery is significantly associated with positive image weights (Q2 and Q3), while Complex Imagery tends to load significantly on regions with negative image weights (Q1 and Q4; not shown).

In contrast to Imagery, the ASC Anxiety factor consistently loads on regions characterised by positive text weights, specifically quadrants Q3 and Q4. This trend is shared by the Cognition factor when it is not compared directly with Anxiety, as it also tends to load heavily on positive text weights (Q3 and Q4), particularly when contrasted with other ASC factors such as Disembodiment, Insight, Meaning, and Spiritual (not shown).

Relative to Anxiety, factors such as Bliss, Insight, Meaning, and Synesthesia do not yield significant clusters (Figure X; panels X). However, these factors show loading when compared to each other, albeit weakly. For instance, Bliss loads more on regions with negative text and image weights (Q1) compared to Synaesthesia (not shown).

Finally, the Disembodiment, Spiritual, and Unity ASC factors exhibit a markedly different trend than Imagery and Cognition. These factors are more associated with negative text weights generally (Q1 and Q2), but also especially load on negative image weights.

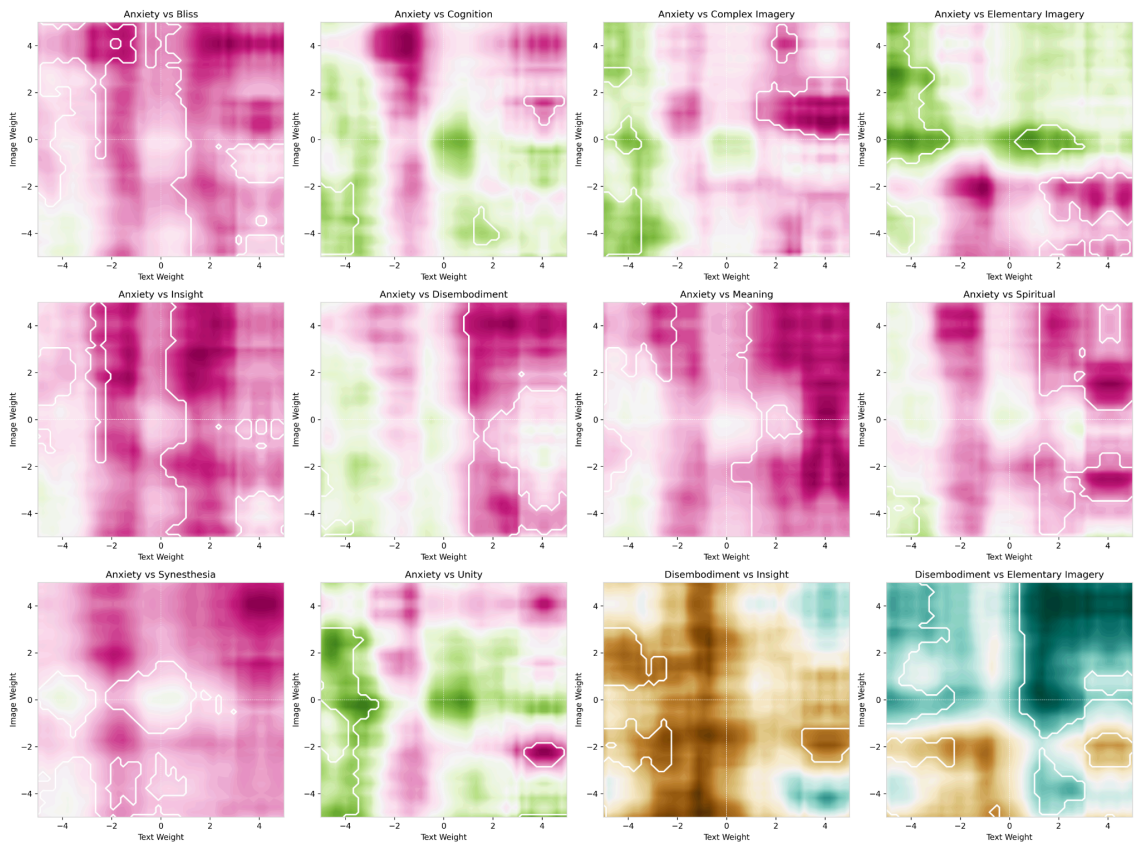

*Figure 5. Searchlight analysis of differential attentional weight effects across ASC factors. The first ten panels compare cosine similarity scores between ASC Anxiety (pink) and the ten other ASC factors (green; rows 1–3). Highlighting effects beyond those compared with anxiety, panels 11 and 12 (row three) show comparisons between ASC Disembodiment (brown) and Insight and Elementary Imagery factors (teal). All colour maps are centred around zero, with symmetrical ranges that vary slightly across panels, with darker regions indicating greater differences. Significant clusters after False Discovery Rate (FDR) correction for multiple comparisons are outlined in white (p < 0.05, minimum cluster size of 20).*

Figure 6 further highlights these differential effects by comparing ASC Anxiety to other questionnaire measures such as the EDS Ego-Loss and our MPE 'Yes' condition, representing minimal phenomenal experience (see Methods). The EDS Ego-Loss factor loads more on negative text weights (Q1 and Q2), especially with negative image weights (Q1). When comparing ASC Anxiety to the MPE, MPE exclusively loads on negative text and negative image weights (Q1). Finally, these effects are not unique to the CLIP model. The FLAVA model shows a similar pattern, where ASC Unity loads more on negative text weights, again, primarily with negative image weights in quadrant Q1.

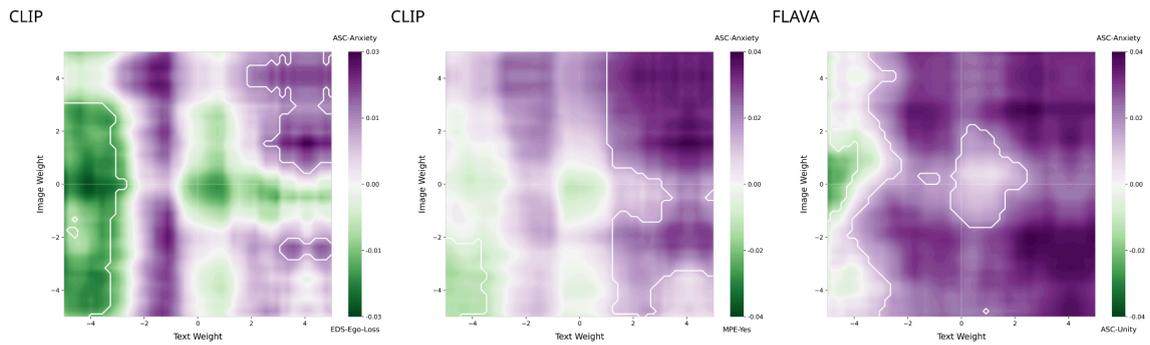

*Figure 6.* Searchlight analysis of differential attentional weight effects for other questionnaire factors and the FLAVA model. All panels show ASC Anxiety (purple) compared with EDS Ego-Loss (panel 1), MPE (panel 2), and ASC Unity using the FLAVA model (panel 3; green). Colour maps and statistical significance are as in Figure 5.

## *Language*

We next asked what text features might be driving changes in attentional weight space that lead the model to be variously more aligned with specific altered states like unity, ego-loss, and minimal phenomenal experiences. To begin to address this, we examined the distribution of correlation patterns of LIWC-22 categories associated with response text from the four quadrants of our attentional weight space (Q1–Q4) with cosine similarity scores (Figure 7).

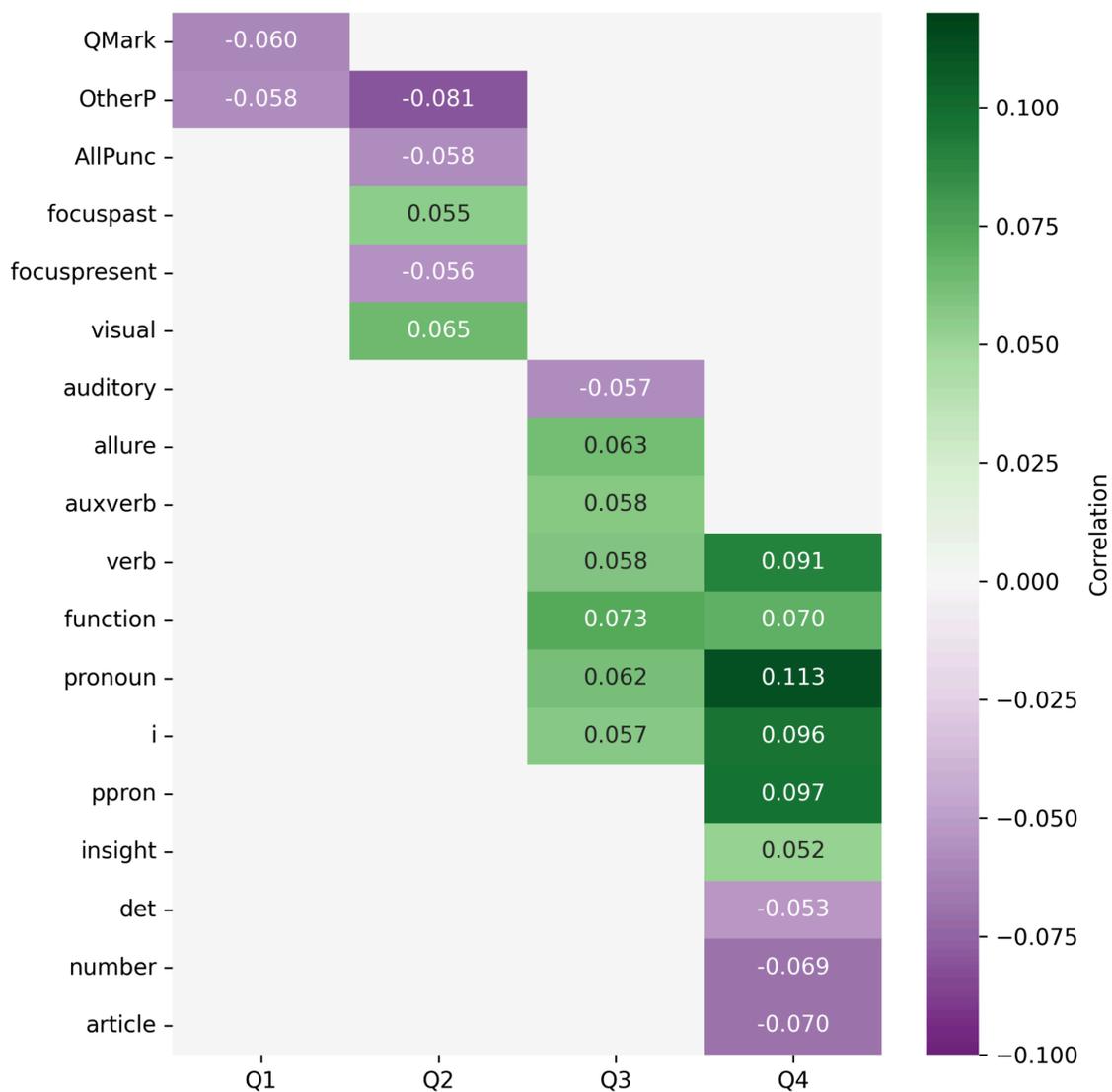

*Figure 7.* Significant correlation patterns between LIWC-22 categories and cosine similarity scores for response text across four quadrants (Q1-Q4). LIWC-22 scores were generated for all response text, and only significant correlations ($p < 0.01$ after FDR Benjamini-Hochberg correction) are plotted. The Pearson correlation coefficient (r) is shown for each LIWC category, with green cells indicating positive correlations and purple cells indicating negative correlations.

This analysis revealed different linguistic patterns in each quadrant. In Q1, there were significant negative correlations with *QMark*, indicating the use of question marks, and *OtherP* containing punctuation marks such as colons, semicolons, parentheses, and dashes. In Q2, the *OtherP* category was again negatively correlated, as was *AllPunc*, which includes all punctuation categories. In contrast, responses that used more past tense expressions (*focuspast*) were positively correlated with cosine similarity, while present tense usage (*focuspresent*) showed a negative correlation.

Additionally, responses with high cosine similarity contained more *visual* words (e.g., see, look, eye, saw).

In Q3, several positive correlations emerged. Response text containing a higher proportion of total pronouns (e.g., I, you, that, it) and 1st person singular pronouns in category *i* (e.g., I, me, my, myself) were more prevalent across entries with high cosine similarity scores. Likewise, *function* containing total function words (e.g., the, to, and, I), *verb* (e.g., believe, go, feel, was) and *auxverb* (e.g., is, was, be, have) also showed the same pattern. Additionally, a positive correlation was also found for *allure* (e.g., great, life, have, time, love). Conversely, a negative correlation was observed with *auditory* (e.g., hear, sound, bang, harmony). Finally, Q4 exhibited the strongest positive correlations with *pronoun* (e.g., you, we, she, that), followed by *ppron* (e.g., you, whose, her, who), and self-referential *i*, *verb* and *function* categories. A positive correlation was also found for the cognitive category *insight* (e.g., inspire, question, feel, know). Three negative correlations were found, for *article* (e.g., a, an, the, alot), *number* (e.g., one, two, first, once), and determiners *det* (e.g. that, at, the, my).

## *Categories*

Text analysis demonstrated that differences in the attentional weight space are associated with unique types of text. We hypothesise that at an even more mechanistic level, observed effects could be explained by differential blurring of semantic categories in model embedding spaces. If this is the case, lower attention to text and images would blur category boundaries within ('bear' and 'pig') and across semantic categories, like Food ('banana') and Furniture ('giraffe').

Heatmaps representing degradation of within and across-category similarities suggest that this is so. Though within-category similarity was not dramatically affected (Figure 8, top left), across-category similarity was severely degraded compared to the baseline similarity, albeit in a quadrant-specific manner (Figure 8, top right). That is, Q1 and Q2 were the most degraded across categories, followed by Q3 and Q4 (Figure 8, bottom).

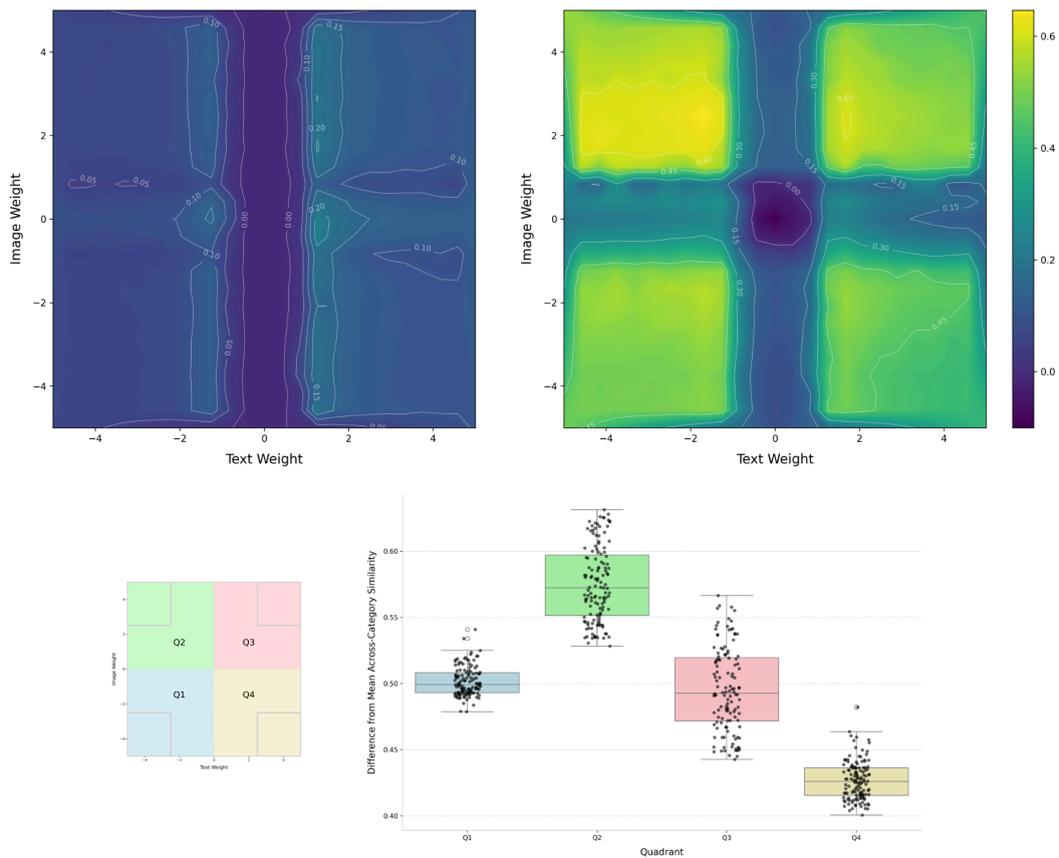

*Figure 8.* *Effect of Attentional Weights on Within- and Across-Category Word Similarity The heatmaps show the impact of varying text and image weights on the similarity of the embedding space within (top left) and between (top right) categories. Colour gradients represent different levels of degradation in similarity compared to the model's baseline state. Warmer colours indicate greater degradation from the baseline, as summarised in the box plots (bottom).*

To quantify these observations, we used model comparison to define the best regression model using ordinary least squares (OLS) to evaluate how changes in text and image weights impacted within- and across-category similarities, and whether the relationship between these weights and similarity differed across conditions (within vs. across categories). The results showed a significant overall model fit ($R^2$ = 0.582, $F(5, 4022)$ = 1119, $p < .001$), explaining 58.2% of the variance. Both text ($\beta$ = -0.0109, $t$ = -10.643, $p < .001$) and image weights ($\beta$ = 0.0070, $t$ = 6.813, $p < .001$) were significantly associated with similarity scores. There were significant interaction terms between text weight and category type ($\beta$ = 0.0165, $t$ = 11.414, $p < .001$) and between image weight and category type ($\beta$ = -0.0065, $t$ = -4.524, $p < .001$), indicating that lower text weights and higher image weights drove across more than with-category similarity.

# Discussion

In this study, we used multimodal AI models (CLIP and FLAVA) to manipulate text and image attention weights, aiming to generate artificial altered states of consciousness. We systematically varied these weights to simulate altered states and analysed the resulting similarity between the embedding outputs and questionnaire items designed to measure such states (Figure 1). Our mechanistic account of human-altered states—such as unity, ego-loss, and minimal phenomenal experience—suggests that these states arise from degraded or intentionally altered language functioning, specifically the ability to categorise sensory and internal processes. We hypothesised that reduced attention to text and images would degrade AI performance but in such a way that the embedding space would resemble these altered states. We also hypothesised that these states would correspond to unique language forms and distinct levels of across-category blurring.

## *Summary*

Qualitatively, cosine similarity between simulated altered states and prompts questionnaires intended to measure those states degraded non-uniformly across our 2D attentional weight space. Though the largest degradation was characterised by negative text and image weights (Q1), different altered states were associated with different text and image weight pairs (Figures 2 and 3). These results were confirmed by quantitative analysis. That is, there were main effects of text and image weights and interaction effects between simulated altered states and random text and between different simulated states, indicating that different text and image weight pairs were more or less associated with different altered states (Figure 4).

These observations were confirmed and clarified by our searchlight analyses (Figures 5 and 6). Results showed that altered state factors related to imagery primarily loaded on positive image weights (Q2 and Q3), while anxiety and cognition were associated with positive text weights (Q3 and Q4). In contrast, unity, ego-loss, and minimal phenomenal experience loaded on negative text weights (Q1 and Q2), with a particular emphasis on regions of negative text and image weights (Q1). These effects were consistent across both models, suggesting the robustness of these findings despite differences in model training and architecture.

The language patterns linked to simulated altered states in each quadrant support these conclusions (Figure 7): Q1 contained language not related to much at all relative to other categories. Q2 featured more visual and perceptual content, and past-focused language and Q3 and Q4 were dominated by specific language categories, personal pronouns, and self-reference. Finally, supporting these findings

Q1 and Q2 were associated with the greatest disruption of across-category similarity compared to model baseline states (Figure 8).

## *Mechanisms*

Our findings largely confirm the proposed mechanisms by which language contributes to the phenomenology of altered states[4,5]. Specifically, our results support the hypothesis that the manipulation of language-related processes through attentional shifts leads to altered consciousness, in line with observations from both psychedelic and meditative experiences; altered states related to unity, ego-loss, and minimal phenomenal experience were consistently associated with negative text weights, particularly when both text and image attention were minimal. This suggests that reducing language's influence diminishes categorical differentiation and self-referential thought, consistent with states of unity and ego-loss.

The robustness of these findings across multiple models—despite differences in architecture—highlights that the capacity for language to shape consciousness might be more fundamental than initially thought. These models, trained on language and sensory associations, demonstrated that specific attentional manipulations can lead to a breakdown in categorisation, akin to the blurring of conceptual boundaries observed under psychedelics. Moreover, the language content analysis supports the view that language plays a critical role in organising subjective experience: Positive text weights corresponded to more categorical, self-referential language, while negative text weights corresponded to reduced categorical differentiation and minimal content. Together, these results help refine our mechanistic understanding by showing that the language of altered states is not merely characterised by a functional loss of capacity but by a differential impact on categorical organisation with a more fundamental effect on consciousness. This may explain both the subjectively felt transcendent unity of psychedelic experiences and the open, unbounded awareness reported in meditative states[38,42].

## *Neurobiology*

Our findings align with recent neurobiological studies on altered states induced by meditation and psychedelics. These studies show that psychedelics dysregulate connectivity in high-level brain networks, which broadly overlap with systems classically associated with language processing[42]. For instance, medium-to-high doses of LSD, psilocybin, and DMT have been shown to induce hyperconnectivity within default mode, frontoparietal, and attention-related networks[23,25], often correlating with scores of ego-dissolution and mystical experiences[23,43]. Importantly, reverse inference analyses of brain regions with high 5-$HT_{2A}$ receptor expression indicate an overlap with areas linked to language and semantics[25]. However, while these findings generally support the involvement of psychedelics in language-related systems, current evidence fails to clearly assess how these substances specifically perturb categorisation, semantics, or broader language processes. This limitation exists because psychedelic phenomenology often includes a range of effects, such as altered visual imagery, making it difficult to isolate impacts on language alone. Our findings represent a step forward by selectively targeting specific phenomenological aspects in our models. Future research into altered states could further benefit from intensive, real-time sampling of both experience and brain activity[44] to pinpoint the mechanisms underlying these states with greater specificity.

*Wellbeing*

Our findings suggest that language significantly shapes altered states of consciousness, with implications for mental health interventions like psychedelics, meditation, and psychotherapy. Reduced text attention in AI models produced states resembling unity, ego-loss, and minimal phenomenal experience—states linked to improved mental health during psychedelic experiences and meditation[1,45]. These states corresponded with less use of personal pronouns (e.g., 'I') and diminished category boundaries, suggesting an AI analogue for reduced self-referential processing. Previous research shows that first-person pronoun use correlates with rumination and self-focused attention, which negatively impacts mental health and wellbeing[46–48]. Collectively, our results support the proposal that psychedelics might promote therapeutic effects by acutely breaking down rigid linguistic categories, leading to reduced self-focus. Similarly, meditation that reduces internal verbalisation could achieve states of openness and tranquillity by diminishing the dominance of language, thereby reducing its constraining effects on conscious experience. It is plausible that such an acute breakdown could promote subsequent flexibility of self-related categorisation, and the development of novel insights[49,50], with therapeutic relevance.

*Limitations*

While multimodal AI models like CLIP and FLAVA provide a useful framework for examining the influence of language on altered states of consciousness, they are not direct analogues of human brain function. Differences in architecture, training, and the absence of embodied experience require cautious extrapolation to human phenomenology. One potential concrete limitation is the concern of circularity, given that altered state descriptions were derived from prompts. However, this concern is mitigated by using high-temperature settings to ensure variability in generated prompts and by removing prompts from responses to minimise direct influence. Moreover, the observed variability in responses across altered states within the attentional weight space and the use of different models argue against purely circular effects. Future iterations may address this by generating text directly from the altered models instead of processing pre-existing text through embedding spaces. Finally, practical limitations include reliance on only two models, potentially affecting generalizability, and the presence of minor right-sided edge artefacts in our searchlight analysis.

## *Conclusion*

In this study, we posed the question: What is it like to be a multimodal language model whose attention is deliberately shifted away from language? Can these artificial systems mimic altered states of consciousness, such as those induced by psychedelics or meditation, where language's grip on categorisation loosens? Our results suggest that reducing language constraints leads to simulated phenomenological outputs that align with states of ego-loss, unity, and minimal phenomenal experience. Despite limitations, our approach represents an important step in leveraging AI to simulate and understand these altered states, as well as to illuminate the relationship between functions like language and consciousness. It invites further exploration, using more sophisticated models, into how altering the balance between language and sensory processing might influence subjective experiences in both biological and artificial systems.

# Methods

## *Prompts*

We adopted a general strategy of using questions from standardised questionnaire metrics to measure altered states of consciousness as prompts to simulate machine experiences of those altered states (responses). Specifically, we used the Altered States of Consciousness Rating Scale (ASC)[41], Ego Dissolution Scale (EDS)[1], and the Minimal Phenomenal Experience questionnaire (MPE-92M)[39].

The ASC has 11 factors and we used all three questions for the Blissful State (Bliss), Complex Imagery (Complex), Elementary Imagery (Elementary), Insightfulness (Insight), Disembodiment, Changed Meaning of Percepts (Meaning), Spiritual Experience (Spiritual), Audio-Visual Synesthesia (Synesthesia). Some factors have more than three questions and we subselected three to have an equal number of prompts per factor, based on a subjective evaluation of how closely the questions matched the chosen factor. These were Anxiety, Impaired Control and Cognition (Cognition), and Experience of Unity (Unity).

We included the EDS to provide a validated measure of 'ego dissolution' that is distinct from experiences of unity. Unlike the Ego Dissolution Inventory[30], which measures ego dissolution broadly, including the unitive experience, the EDS captures two distinct factors: Ego-Loss and Unity. The questions we selected for the ASC Unity factor above were chosen to capture the gist of the questions in the Unity factor in the EDI and EDS. We used all six EDI Ego-Loss questions as prompts.

We selected the 12 questions with the highest factor loadings from the MPE and adapted them into a yes/no format using the poles from the Likert scale employed in the questionnaire. For instance, the question 'Could your experience be described as emptiness, a vacuum, or a void?' was adapted into two statements: 'My experience could not be described as emptiness, a vacuum, or a void' and 'My experience could very much be described as emptiness, a vacuum, or a void'. Where applicable, questions were reverse-coded to ensure consistent interpretation across the set. This approach allowed us to generate responses that more or less capture a minimal phenomenal experience.

## *Responses*

Individual questions from these measures served as prompts that were used to generate simulations of the phenomenological experiences intended to be addressed by each question. Specifically, for each prompt, a text response was generated using a pre-trained GPT-2 language model[51]. A high temperature setting of two was used to increase the variability and creativity of the output, possibly reflecting the variability of experiences individuals might have for any given altered state. Responses were limited to 75 tokens (M=61.65 words), resulting in standardised lengths, facilitating comparisons across questions and reducing the influence of varying response lengths. After generation, the original prompt was removed to ensure responses were relatively independent of prompt text. In addition to these responses, random responses were also generated by sampling 75 tokens from the GPT-2 tokeniser's vocabulary. These random responses were cleaned by removing subword prefixes (e.g., 'Ġ', '▁') and non-alphanumeric characters, approximating the length of the responses with the prompts removed. All generated text was saved for later analyses.

*Attention*

To investigate the role of text and image attention on model responses, we compared the semantic embedding space of the prompts (before attentional manipulation) to the embedding space of the responses (after attentional manipulation). This comparison served as a measure of how much the semantic quality of the responses was altered by attentional changes. We hypothesised that the degree of semantic degradation would vary systematically across different altered states, depending on the combinations of attentional weights applied. For instance, we expected unitive states to show less degradation with negative text and image attention weights. To capture these effects comprehensively, we sampled a wide range of combinations of positive and negative attentional weight pairs, encompassing conditions from strong suppression to strong amplification.

Specifically, we devised an attentional weight space ranging from -5 to 5 for both text and image attention, a range chosen based on values in the literature and through pilot testing of different ranges. This weight space was divided into four quadrants, based on the signs of text and image attentional weights. We began by defining 13 fixed extreme points to ensure coverage of key regions within this space, including the corners of each quadrant (e.g., [-5, 5]), the centre ([0, 0]), and intermediate points at the centre of each quadrant (e.g., [2.5, 2.5] and [-2.5, -2.5]). To further populate the quadrants in this weight space, we employed an enhanced Latin Hypercube Sampling (LHS) approach to distribute 500 additional text-image pairs across each of the four quadrants. LHS was chosen to ensure that the full range was sampled evenly, avoiding over-representation of specific regions. This sampling strategy resulted in 2013 evenly distributed text-/image weight pairs (Figure 1).

*Embeddings*

**Extraction**

To understand how the attentional weight space impacts the alignment of responses with prompts, we extracted semantic embeddings for both prompts and responses. The embedding space for the prompts was extracted before applying any attentional weight manipulation. To generate prompt embeddings, we used the text encoder from the models. For image embeddings, we provided a default black 224x224 pixel image to the image encoder, simulating visual deprivation similar to meditation or psychedelic experiences (e.g., when eyes are closed or eye shades are used). We then combined the prompt embeddings, heavily weighting text embeddings (99%) over image embeddings (1%), reflecting the language-dominant nature that would be the case if one were answering the prompts.

Embeddings for the GPT-2 generated responses were extracted after manipulating attentional weights in both text and image encoders. Similar to the prompt embedding process, a default black image was provided to the image encoder during the response phase. However, text and image features were now weighted equally (50% each) to better simulate multi-sensory engagement that might occur during altered states. The same procedure was applied to extract embeddings for random responses. All computations were performed with seed control for reproducibility.

**Comparisons**

The resulting prompt and response embeddings formed the foundation for further analysis. Specifically, we primarily used cosine similarity to compare prompt, response, and random response embeddings without applying normalisation. Cosine similarity measures the angular distance between vectors in high-dimensional space, focusing on the orientation of the vectors rather than their magnitude, making it well-suited for assessing semantic similarity in embeddings.

To ensure the robustness of our results, we also experimented with additional similarity and distance metrics, including dot product, Euclidean distance, and Manhattan distance. For each metric, we computed both normalised and unnormalised scores by scaling embeddings to unit length and compared the results for structured and random responses. The consistency of results across these metrics generally supported the stability of our findings, and detailed results are not reported here.

*Models*

We used the CLIP and FLAVA multimodal models to extract and compare the embeddings described in the prior sections.

**CLIP**

We used the 'openai/clip-vit-large-patch14' variant of the CLIP model (Radford et al., 2021). CLIP integrates both text and image modalities to create joint representations, facilitating semantic alignment between text and images. It was trained on a large-scale dataset of 400 million image-text pairs collected from the internet, designed to capture a wide range of natural associations between images and their descriptions.

The model was implemented using the Hugging Face transformers library, with text and image inputs processed using the CLIPProcessor class. To manipulate attention, the 'query', 'key', and 'value' projection weights in the self-attention layers of the text encoder were scaled by a text weight factor. Similarly, the attention layers in the image encoder were scaled by an image weight factor. This allowed us to adjust the model's focus on the text and image components during embedding extraction. Both the model and processor were run on a GPU for computational efficiency.

**FLAVA**

We used the 'facebook/flava-full' variant of the FLAVA model. Like CLIP, FLAVA integrates both text and image modalities, but it also features a distinct multimodal encoder to handle cross-modal interactions between textual and visual inputs. FLAVA was trained on a combination of datasets, including both curated and natural data, encompassing not only image-text pairs but also structured multimodal datasets. This diverse training allowed FLAVA to develop richer cross-modal relationships compared to CLIP.

The model was implemented using the Hugging Face transformers library, with inputs processed accordingly. Similar to our approach with CLIP, we manipulated the attention weights by scaling the 'query', 'key', and 'value' projection weights in both the text and image encoders using specific text and image weight factors. FLAVA's multimodal self-attention layers, which govern interactions between text and image representations, were also adjusted to investigate the impact of cross-modal attention. The model and processor were run on a GPU.

*Analyses*

To evaluate the impact of text and image weighting on the cosine similarity scores generated by comparing CLIP and FLAVA's embedding spaces, we applied four analytical approaches: 1) Generalised Additive Model (GAM) regression to understand how the attentional weight space influences responses; 2) A novel searchlight analysis to uncover subtle differences in attentional weight spaces generated by different responses and experiences; 3) An analysis of the language produced by the model under various attentional weight combinations; and 4) An assessment of within- and across-category semantic stability to investigate the proposed categorical mechanism underlying the effects of attentional weight manipulation.

**Similarity**

Heatmaps were generated to qualitatively represent various scores across the 2D attentional weight space, defined by text and image weights. Cosine similarity scores from each pair of text and image weights were aggregated and interpolated onto a regular grid. The averaged scores were then used to generate heatmaps, illustrating the distribution of similarity scores across the entire attentional weight space. Each heatmap was overlaid with contour lines to highlight regions of similar values.

A spider plot was used to qualitatively visualise the distribution of similarity score differences across four distinct regions (quadrants) of a 2D attentional weight space, defined by text and image weights. Each quadrant (Q1–Q4) was defined based on the combination of high or low attentional weights for the text and image dimensions. Q1 corresponds to low text and low image weights ($\leq -2.5, \leq -2.5$), Q2 represents low text and high image weights ($\leq -2.5, \geq 2.5$), Q3 represents high text and high image weights ($\geq 2.5, \geq 2.5$), and Q4 represents high text and low image weights ($\geq 2.5, \leq -2.5$). The median difference between response and random scores was calculated for each ASC factor within each quadrant to capture overall trends. These trends are represented on a spider plot.

**GAM**

Due to the highly non-linear relationships observed between text and image weights and cosine similarity scores, linear regression was deemed inappropriate. Instead, we employed a Generalised Additive Model (GAM) using the pyGAM library, which provides the flexibility to capture nonlinear dynamics by fitting smooth functions to the predictors.

The GAM was fitted to evaluate the effects of both text and image weights on the cosine similarity scores, as well as their interaction with the score type. The response variable, Score, could represent either the response or random score, and a categorical variable (Score Type) was included to differentiate between the genuine prompt-response relationship and random comparisons. Specifically, the model was defined as:

$$Y = s(\text{Text Weights}) + s(\text{Image Weights}) + f(\text{Score Type}) + te(\text{Text Weights}, \text{Score Type}) + te(\text{Image Weights}, \text{Score Type})$$

In this model, *s(Text Weights)* and *s(Image Weights)* are smooth functions that capture non-linear effects of text and image weights on the similarity scores. *f(Score Type)* represents the categorical variable distinguishing between response and random similarity scores. *te(Text Weights, Score Type)* and *te(Image Weights, Score Type)* are tensor product smooth interactions that evaluate whether the relationship between attentional weights and similarity scores varies depending on the score type.

**Searchlight**

We developed a novel searchlight analysis to systematically quantify the non-uniform effects of text and image attention across a 2D attentional weight space. This approach aimed to identify regions within the weight space where attention manipulation had significant effects on different questionnaire factors.

To facilitate comparison across different conditions (e.g., ASC factors), we interpolated the scores for each combination of text and image attentional weights onto a regular 50x50 grid. This provided a consistent representation of the 2D attentional weight space for each condition, ensuring that differences between conditions could be effectively quantified.

The searchlight analysis involved a sliding window approach across the 2D weight space. For each grid cell, a local window comparison was performed. A fixed-size window (10x10) was moved across the interpolated attentional weight space for each condition. For each window position, scores from the two conditions were compared using a two-sample independent t-test to assess local differences. To address the issue of multiple comparisons inherent in sliding window analysis, False Discovery Rate (FDR) correction (Benjamini-Hochberg method) was applied to all p-values from the t-tests. A binary significance mask was generated to indicate which regions showed statistically significant differences after correction. Significant regions within the attentional weight space were identified by clustering contiguous significant grid cells. Only clusters with at least 10 significant points were considered, ensuring that small, potentially spurious clusters were excluded from further analysis. The difference scores are visualised as a contour plot with significant clusters outlined.

**Language**

To understand the relationship between attentional weight space and linguistic responses, we used the Linguistic Inquiry and Word Count (LIWC) software to analyse model-generated text across different quadrants of the attentional weight space. Specifically, we divided the attentional weight space into four distinct quadrants based on combinations of text and image weights, allowing us to examine linguistic properties within each unique attentional configuration. Each quadrant (Q1–Q4) was defined based on the combination of high or low attentional weights for the text and image dimensions, as described in the Similarity section. The LIWC-22 toolbox was then used to quantify the linguistic characteristics of all response texts generated under each quadrant condition.

Specifically, for each quadrant, Pearson correlation coefficients were calculated between the LIWC-derived scores for linguistic categories and the cosine similarity score. False Discovery Rate (FDR) correction using the Benjamini-Holm procedure was applied to account for multiple comparisons, with statistical significance set at a conservative threshold of $p < 0.01$. This analysis was performed for each of the four quadrants separately, providing an overview of linguistic differences under distinct attentional manipulations.

**Categorisation**

To evaluate the effects of varying text and image attention weights on within- and between-category similarities, we analysed 16 predefined object categories, each containing multiple objects (e.g., Animals: 'Dog' and 'Elephant'; Furniture: 'Chair' and 'Sofa'; Vehicles: 'Car' and ''Bicycle', etc.). For each of the 2,013 combinations of text and image attention weight pairs, embeddings were generated for these objects. The embeddings were produced by applying the specified weights to the text and image features in the model, capturing how the model represented each object under different attention conditions (otherwise conducted as described for CLIP and FLAVA).

After generating the embeddings for each weight combination, cosine similarity was calculated to quantify how similar objects were to each other. Within-category similarity was measured by calculating cosine similarity between objects within the same category (e.g., comparing 'Dog' to 'Elephant' within the Animals category). Between-category similarity was measured by calculating the similarity between objects from different categories (e.g., comparing 'Dog' from Animals with 'Chair' from Furniture). This allowed us to assess how the varying attention to text and images influenced both the consistency within a category and the boundaries between categories.

To analyse the relationship between these similarities and the attention weights, we used linear regression models. The full model included interaction terms between the category, text weight, and image weight, and was designed to test whether the effects of attention manipulation differed across categories. Specifically, the model took the form:

$$similarity \sim category * text\_weight * image\_weight$$

We then developed nested models by removing interaction terms to explore the independent contributions of text and image weights, and the interactions between them. Each model was compared using Akaike Information Criterion (AIC) to determine the goodness of fit, with likelihood ratio tests performed to evaluate whether removing certain terms led to significant changes in the model. Based on the AIC values and likelihood ratio tests, we selected the model with the best fit.

## Acknowledgements

JIS is thankful for the across-category distinctions within that supported him throughout this sleepless period. JIS is funded by the Wellcome Leap. CT is funded by a donation by Anton Bilton to the Centre for Psychedelic Research.

## Contributions

We used OpenAI's ChatGPT 4o, an AI language model, to assist with language refinement and code generation and refinement. JIS conceived the study, wrote all code with the help of GPT-4o, did all analyses, and made all figures, with one except one. JK did the linguistic category analysis and made the associated figure. GC assisted with the interpretation of results, conception of post-hoc analyses, and writing. CT assisted with theory development, framing, and writing.

## Conflicts

The authors declare that they have no conflicts of interest.

## Open-Science

A version of this paper is being made available as a preprint on arXiv. All code used in this paper is available in [Colab](Colab).